\documentclass[twocolumn,amsmath,amssymb,superscriptaddress,aps,prb]{revtex4-1}
\usepackage{stackengine,graphicx} 
\usepackage{epsfig} 
\usepackage{dcolumn} 
\usepackage{bm} 
\usepackage{physics}
\usepackage{amsmath}
\usepackage{xcolor}
\usepackage{xspace}
\usepackage{outlines}
\usepackage{textcomp}
\usepackage{gensymb}
\usepackage[normalem]{ulem}
\graphicspath{{./}}
\newcounter{saveenumi}

\newcommand{\be}{\begin{enumerate}}
\newcommand{\ee}{\end{enumerate}}

\definecolor{robgreen}{HTML}{00a832}
\definecolor{robadd}{HTML}{005c12}
\definecolor{emiladd}{HTML}{1438c7}
\definecolor{black}{HTML}{000000}
\definecolor{linebrown}{HTML}{bd6e00}
\definecolor{lineblue}{HTML}{0000ff}
\definecolor{linegreen}{HTML}{217200}
\definecolor{linered}{HTML}{d20d0d}

\def\rjkadd#1{\textcolor{black}{#1}}

\newcommand{\RAA}{\AA$^{-1}$}

\def\rup{RuP}
\def\fd3m{Fd$\overline 3$m}
\def\i41amd{I4$_{1}$/amd}
\def\t2g{$t_{2g}$}

\newcommand{\ia}{\ensuremath{\mathrm{\AA}^{-1}}\xspace}
\newcommand{\qmax}{\ensuremath{Q_{\mathrm{max}}}\xspace}

\newcommand{\pdfgui}{\textsc{PDFgui}\xspace}

\newcommand{\pdfgetxthree}{\textsc{PDFgetX3}\xspace}

\newcommand{\pyfai}{\textsc{pyFAI}\xspace}

\newcommand{\cmi}{\textsc{DiffPy-CMI}\xspace}

\newcommand{\soutoldold}[1]{}
\newcommand{\soutold}[1]{}

\begin{document}
%
%
\begin{abstract}
Superconductivity in binary ruthenium pnictides occurs proximal to and upon suppression of a mysterious non-magnetic ground state, preceded by a pseudogap phase associated with Fermi surface instability, and its critical temperature, $T_{c}$, is maximized around the pseudogap quantum critical point.
By analogy with isoelectronic iron based counterparts, antiferromagnetic fluctuations became ''usual suspects" as putative mediators of superconducting pairing.
Here we report on a high temperature local symmetry breaking in RuP, the parent of the maximum-$T_{c}$ branch of these novel superconductors, revealed by combined nanostructure-sensitive powder and single crystal X-ray total scattering experiments.
Large local Ru$_{6}$ hexamer distortions associated with orbital-charge trimerization form above the two-stage electronic transition in RuP.
While hexamer ordering enables the nonmagnetic ground state and presumed complex oligomerization, the relevance of pseudogap fluctuations for superconductivity emerges as a distinct prospect.
As a transition metal system in which partial d-manifold filling combined with high crystal symmetry promotes electronic instabilities, this represents a further example of local electronic precursors underpinning the macroscopic collective behavior of quantum materials.
\end{abstract}

\title{Role of Local Ru Hexamers in Superconductivity of Ruthenium Phosphide}

\author{Robert~J.~Koch}
\affiliation{Condensed Matter Physics and Materials Science Division, Brookhaven National Laboratory, Upton, NY 11973, USA}
\author{Niraj~Aryal}
\affiliation{Condensed Matter Physics and Materials Science Division, Brookhaven National Laboratory, Upton, NY 11973, USA}
\author{Oleh~Ivashko}
\affiliation{Deutsches Elektronen-Synchrotron DESY, Notkestrasse 85, Hamburg, D-22607, Germany}
\author{Yu~Liu}
\affiliation{Condensed Matter Physics and Materials Science Division, Brookhaven National Laboratory, Upton, NY 11973, USA}
\affiliation{Los Alamos National Laboratory, Los Alamos, New Mexico 87545, USA}
\author{Milinda~Abeykoon}
\affiliation{Photon Sciences Division, Brookhaven National Laboratory, Upton, NY 11973, USA}
\author{Eric~D.~Bauer}
\affiliation{Los Alamos National Laboratory, Los Alamos, New Mexico 87545, USA}
\author{Martin~v.~Zimmermann}
\affiliation{Deutsches Elektronen-Synchrotron DESY, Notkestrasse 85, Hamburg, D-22607, Germany}
\author{Weiguo~Yin}
\affiliation{Condensed Matter Physics and Materials Science Division, Brookhaven National Laboratory, Upton, NY 11973, USA}
\author{Cedomir~Petrovic}
\affiliation{Condensed Matter Physics and Materials Science Division, Brookhaven National Laboratory, Upton, NY 11973, USA}
\author{Emil~S.~Bozin}\email[]{bozin@bnl.gov}
\affiliation{Condensed Matter Physics and Materials Science Division, Brookhaven National Laboratory, Upton, NY 11973, USA}

\maketitle

\section{Introduction}
\label{section:intro}

Spatiotemporal fluctuations of electronic charge, orbital, and spin are native to the quantum materials realm~\cite{keime;np17,bramw;prl00,boyar;ltp05,fernan;arcmp19,pasch;nrp21},
encompassing phenomena ranging from unconventional superconductivity~\cite{auvra;nc19,chen;nm19}, charge density waves~\cite{torch;nm13,joe;np14}, and pseudogaps~\cite{rice;ssc73} to metal-insulator transitions~\cite{hosho;prl17,feng;nc21}, colossal magnetoresistivity~\cite{sheno;prl07}, and frustrated magnetism~\cite{stock;prl09,freel;npjqm21}.
Elucidating the character and role of electronic short-range correlations in the emergence of application-relevant phases~\cite{watan;jpsj02,yokoy;prb05,drew;prl08,glasb;np14,leder;prl15,frand;prl17,loret;np19,arpai;s19,poude;npjqm19,wu;prl21} is challenging as most experimental probes yield  bulk averages~\cite{hoshi;prl15,ikeda;pnas21,schaf;jpcm21}.
When the crystal lattice is involved, the pair distribution function (PDF) approach provides a unique perspective informing on the presence and nature of states of local broken symmetry on the nanoscale~\cite{billi;s07}.
This is exemplified by the recent observation of structural fluctuations appearing alongside ferromagnetic order~\cite{perve;nc19} greatly above the Verwey transition in magnetite (Fe$_{3}$O$_{4}$)~\cite{verwe;n39}, which illuminated a century old mystery of its mechanism~\cite{walz;jpcm02} by revealing the essential role of magnetism in formation of the trimeron ground state~\cite{senn;n12}.
Similarly, the discovery of structural fluctuations in iridium thiospinel (CuIr$_{2}$S$_{4}$)~\cite{bozin;nc19} unmasked an orbital precursor to its metal-insulator transition~\cite{furub;jpsj94}, amending the orbital-selective mechanism of the octamer molecular orbital crystal state formation~\cite{radae;n02} proposed based on crystallography alone~\cite{khoms;prl05}.

Here we report on preformed local hexamer distortions above the two-stage electronic phase transition in RuP detected by combined X-ray total scattering based powder and single crystal PDF analyses.
Metallic orthorhombic ($Pnma$) binary ruthenium pnictides Ru$Pn$ ($Pn$=P, As, Sb), isoelectronic with iron pnictides~\cite{si;nrm16}, were first explored for superconductivity (SC) by Hirai et al~\cite{hirai;prb12}.
While RuSb displays SC below 1.2~K, RuP and RuAs have a non-magnetic (NM) ground state reached via two successive temperature driven electronic transitions, at T$_{1}$ and T$_{2}$ (T$_{2} <$~T$_{1}$), accompanied by global symmetry lowering~\cite{hirai;prb12,chen;prb15,koteg;prm18}, and SC emerges only upon electron doping.
Continuous transition at T$_{1}$ to an intermediate pseudogap (PG) phase~\cite{hirai;prb12,sato;arxiv12,li;prb17} is followed by a first order transition to an insulating state at T$_{2}$ where a sharp drop in susceptibility exposes core diamagnetism~\cite{hirai;prb12} consistent with spin-singlet formation~\cite{koteg;prm18,kuwat;jpsj18}.
Substitution of Rh for Ru rapidly suppresses the NM phase, while the PG phase persists with eventual abrupt collapse at much higher doping.
A familiar composition-temperature phase diagram~\cite{basov;np11,husse;rpp18} ensues: a dome of superconductivity appears in a narrow composition range around quantum critical point for the PG phase at which the SC temperature displays a maximum (3.7~K in Ru$_{1-x}$Rh$_{x}$P and 1.8~K in Ru$_{1-x}$Rh$_{x}$As).
Stereotypically, this landscape suggests that fluctuations of some order parameter may aid superconductivity in Ru$Pn$.
Antiferromagnetic fluctuations were seen in the PG regime~\cite{li;prb17,kuwat;jpsj18}, growing on approaching the NM state and weakening appreciably towards the SC compositions~\cite{li;prb17,kuwat;jpsj18}.
Charge and orbital sectors~\cite{sprau;s17} have not been explored in this context despite anomalous metallicity above T$_{1}$ in RuP and RuAs, where a rather broad Drude-like component in optical conductivity was observed~\cite{chen;prb15,nakaj;prb19}, indicative of strong carrier scattering.

Various mechanisms for the NM ground state have been considered, including charge density wave, spin-singlet, orbitally driven Peierls, and valence bond crystal featuring dimers, polymers, or molecular chains~\cite{ootsu;prb20}, but the driving force behind the two-step transition remains elusive, partially due to lack of complete crystallographic information.
Below T$_{2}$ RuAs becomes globally monoclinic ($P2_1/c$), but neither dimerization nor trimerization features were identified~\cite{koteg;prm18}.
The structures of RuAs at T$_{2} <$ T $<$ T$_{1}$ and RuP below T$_{1}$ remain undetermined to date.
The instability of the electronic structure imposed by global $Pnma$ symmetry of RuP and RuAs, exhibiting degenerate narrow flat bands derived chiefly from Ru 4$d_{xy}$ orbitals with corresponding partial density of states (DOS) sharply peaked at E$_{f}$~\cite{goto;pp15,koteg;prm18}, is presumed to be removed below T$_{2}$ and to be the root cause of the two transitions~\cite{ootsu;prb20}.
The PDF analysis of RuP unveils that the symmetry associated with this instability is actually removed already in the metallic (T~$>$~T$_{1}$) regime via local symmetry breaking, which not only acts as a precursor to the NM ground state, but may also be relevant for superconductivity in binary ruthenium pnictides.

\section{Results}
\label{section:res}
\subsection{Average structure and properties}
%

\begin{figure}[tb]
\includegraphics[width=0.485\textwidth]{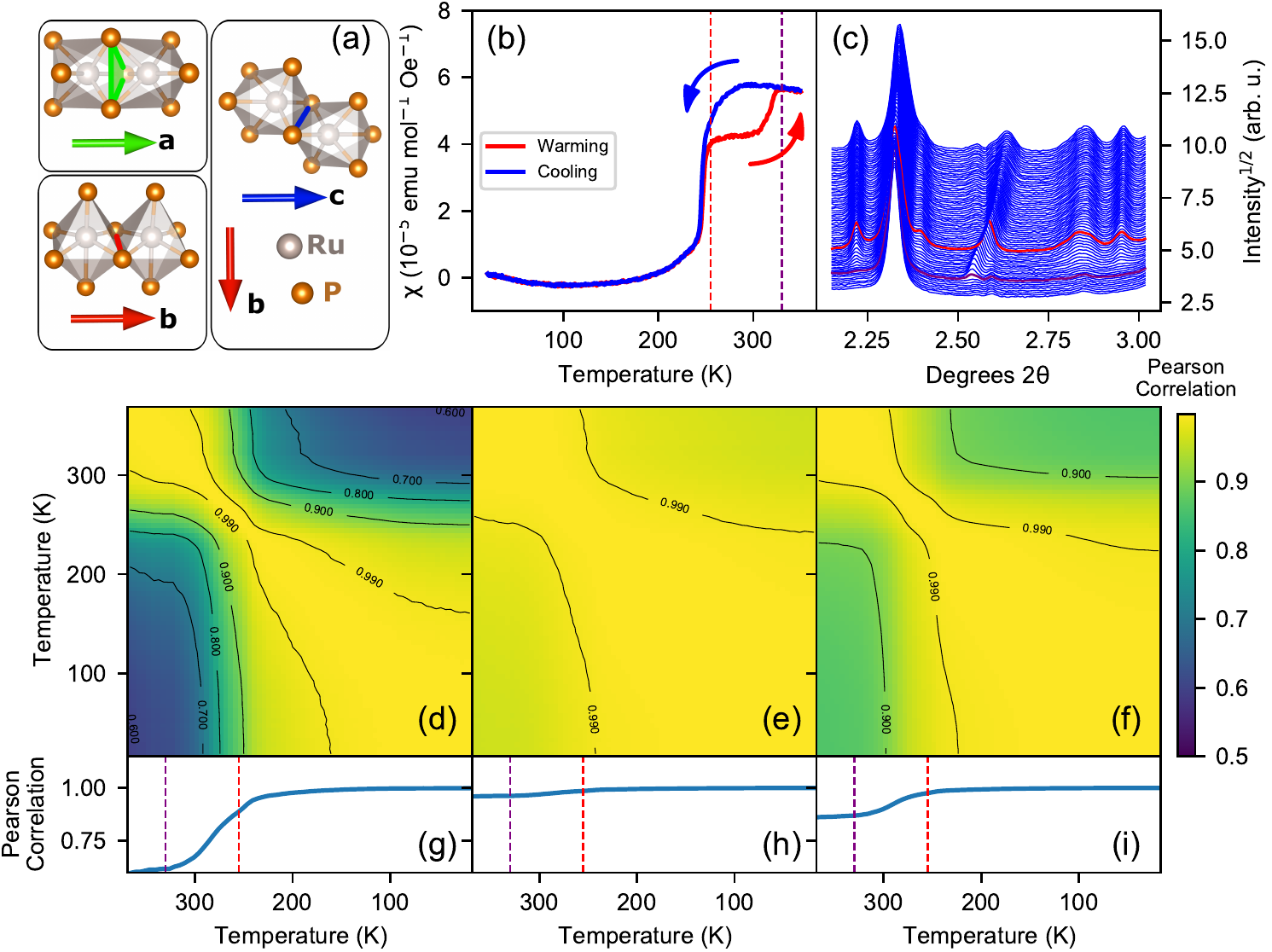}
\caption{\label{fig:struc_prop}
RuP properties and transitions.
(a) Three distinct types of RuP$_6$ octahedra connectivity: face sharing along the $a$-axis (green contacts), and edge sharing along the $b$- and $b/c$-axes (red and blue contacts, respectively).
(b) Magnetic susceptibility (single crystal) on cooling and warming, depicting two transitions at T$_1$ = 330~K (purple dashed line) and T$_2$ = 260~K (red dashed line).
(c) Temperature dependent powder x-ray diffraction data (narrow $2\theta$ range), with temperature decreasing from bottom to top, from 370~K to 10~K in 5~K steps.
The diffraction pattern for T$_1$ = 330~K and T$_2$ = 260~K are shown in purple and red, respectively.
Pearson correlation coefficient, computed between distinct temperature points comparing the PDF signal across (d ,g) 25-60~\AA\ or (e, h) 0-10~\AA, as well as comparing (f, i) the powder XRD signal across 1-15 degrees $2\theta$.
Panels (d, e, f) show the full Pearson correlation maps, while panels (g, h, i) present line cuts of these maps for T = 10~K, with T$_1$ and T$_2$ marked by purple and red dashed lines, respectively.
}
\end{figure}

\begin{figure}[tb]
\includegraphics[width=0.485\textwidth]{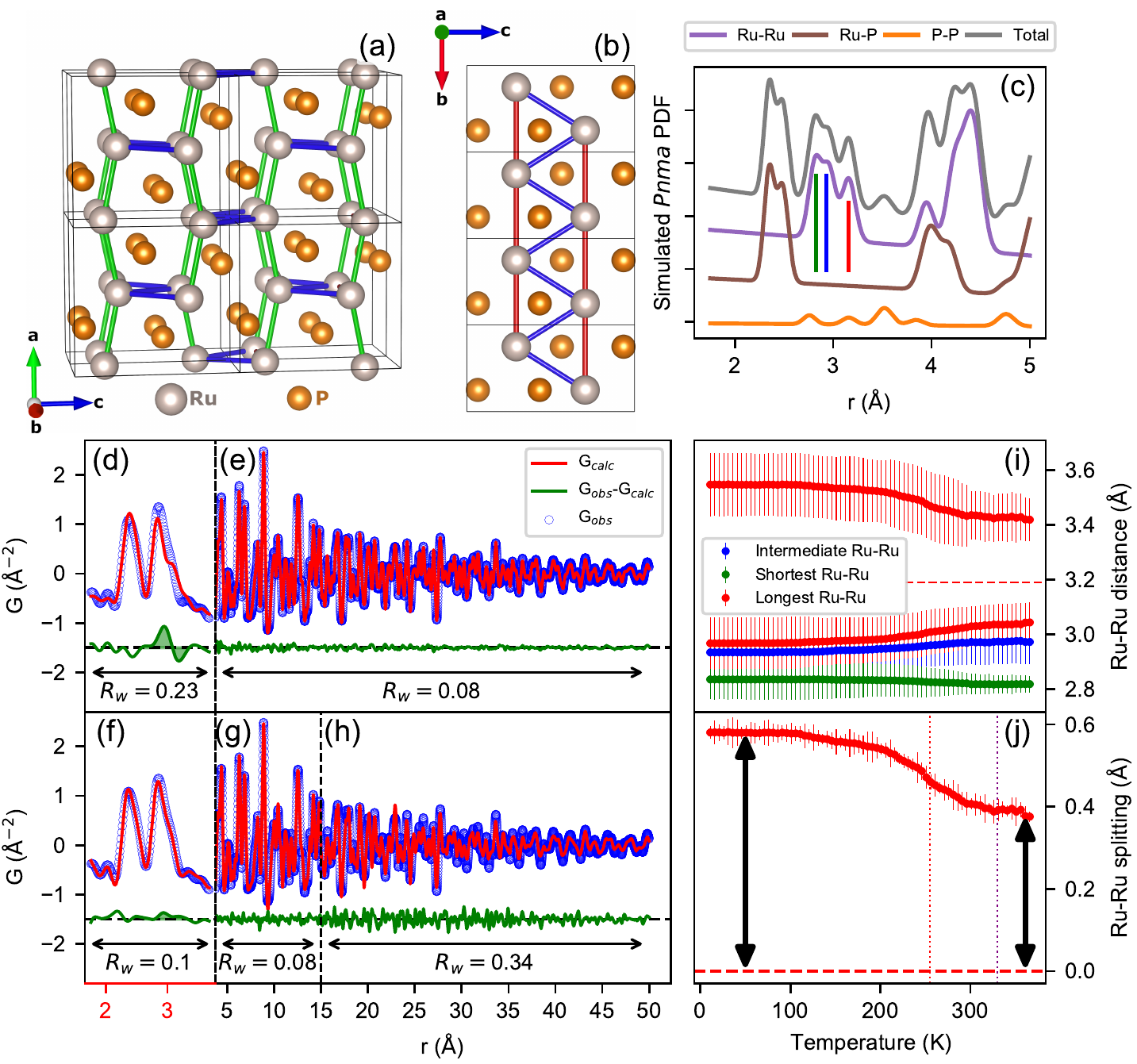}
\caption{\label{fig:misfit}
Local structure from powder PDF.
Eight unit cells of the $Pnma$ structure of \rup\ viewed along the (a) $b$-axis and (b) a-axis.
Three distinct Ru-Ru nearest neighbor pairs are shown as solid lines: green, blue, and red contacts correspond to the shortest, intermediate, and longest pair distances, respectively.
(c) A simulated \rup\ x-ray PDF, decomposed into the partial pair contributions.
The PDF contributions from the three $Pnma$ symmetry-allowed Ru-Ru nearest neighbor pairs shown in (a,b) are identified by identically colored vertical lines.
PDF fit of the experimental PDF signal measured at 370~K utilizing the $Pnma$ model shows (d) a large misfit at ca. 3.1~\AA, the approximate location of Ru-Ru pairs.
The fit quality over this range is poor, resulting in  $R_w = 23\%$.
Over (e) a broader $r$-range, the $Pnma$ model fit is adequate, with $R_w = 8\%$.
PDF fit of the 370~K data utilizing the $P2_1/c$ constrained model (f) remedies the misfit at ca. 3.1~\AA, resulting in $R_w = 10\%$.
The fit is adequate (g) up to about 15~\AA, with $R_w = 8\%$, but (h) extending the calculation range reveals inadequacy of local structure model in describing the longer range structure ($R_w = 34\%$).
(i) Ru-Ru pair distance vs. temperature, extracted from the $P2_1/c$ fit up to 15~\AA\ reveals the primary symmetry breaking on the longest Ru-Ru pairs, parallel to the $Pnma$ $b$-axis.
(j) The magnitude of the Ru-Ru distance splitting along the $Pnma$ $b$-axis.
}
\end{figure}

For T~$> 330$~K RuP adopts the MnP-type $Pnma$ structure~\cite{hirai;prb12}, featuring RuP$_6$ octahedra with face-sharing along the $a$-axis and edge sharing along the $b$-axis and in the $bc$ plane (Fig.~\ref{fig:struc_prop}(a)), creating 3 nonequivalent Ru-Ru nearest neighbor (NN) pairs. Shown in Fig.~\ref{fig:misfit}(a, b),
the shortest contacts (green) constitute zig-zag chains along the $a$-axis, whereas the intermediate (blue) and long (red) contacts form zig-zag ladders in the $bc$ plane comprised of straight rails (red) and zig-zag rungs (blue).

Magnetic susceptibility, after subtraction of a low-temperature Curie tail (Fig.~\ref{fig:struc_prop}(b)), consistent with a nonmagnetic ground state evidences two electronic phase transitions, seen
by total scattering as two global symmetry breaking structural transitions.
Formation of new, weak, Bragg peaks at $T_{1} = 330$~K (Fig.~\ref{fig:struc_prop}(c), purple trace) in the powder diffraction data, specifically at ca. 2.20 and 2.55 degrees $2\mathrm{\theta}$, indicates an onset of ordering below the PG transition temperature.
Further nontrivial changes in the diffraction profile around $T_{2} = 260$~K (Fig.~\ref{fig:struc_prop}(c), red trace) confirm that both transitions take place in the sample.

The 370~K Bragg data are consistent with the $Pnma$ symmetry (Rietveld fit, Supplementary Note 1).
However, the model yields rather large U$_{22}$ atomic displacement parameter (ADP) of Ru (0.0168(5)~\AA$^{2}$) as compared to U$_{11}$ (0.0042(5)~\AA$^{2}$) and U$_{33}$ (0.0054(3)~\AA$^{2}$).
This adds to the list of high temperature anomalies and points to structural disorder along the $b$-axis.

The Pearson correlation coefficient (defined in Methods) calculated for temperature dependent powder PDF data over an $r$-range which includes only the long-range structure and excludes the local structure ($15<r<60$~\AA, Fig.~\ref{fig:struc_prop}(d, g)), shows two distinct regions of self-similarity, with a single region of strong dissimilarity bounded by $T = 260$~K, demonstrating that a long-range structural phase transition at T$_{2}$ is apparent here.
Conversely, when considering an $r$-range which includes only the local structure ($0<r<15$~\AA, Fig.~\ref{fig:struc_prop}(e, h)), the region of dissimilarity at $T = 260$~K is considerably less pronounced, with no discernible change at T$_{1}$.
The Pearson correlation coefficient of the total scattering signal (Fig.~\ref{fig:struc_prop}(f, i)), which represents the effects of both the long-range and local structure, again shows a single region of dissimilarity bounded by $T = 260$~K.
This dissimilarity is weaker than that observed when considering the long-range PDF (Fig.~\ref{fig:struc_prop}(d, g)) but stronger than that observed when considering the local structure PDF (Fig.~\ref{fig:struc_prop}(e, h)).
This correlation analysis establishes distinct manifestations of \rup\ structure on different length-scales.

\subsection{Local structure from powder PDF}

To elucidate the nature of structural disorder above the PG phase (T~$>$~T$_{1}$) and its evolution on cooling, we turn to powder PDF analysis.
A fit of the $Pnma$ symmetry model to 370~K data over $1.75<r<50$~\AA\ range, shown in Fig.~\ref{fig:misfit}(d, e), results in an overall fit residual of $R_w = 11\%$.
However, while the model adequately reproduces the observed PDF for $r>3.9$~\AA, where $R_w = 8\%$, it fails for $r<3.9$~\AA, where $R_w = 23\%$.
A large misfit at ca. 3.1~\AA, highlighted on residual curve in Fig.~\ref{fig:misfit}(d), reveals a short-range distortion incompatible with average symmetry.
The $Pnma$ model, which is consistent with the majority of the observed PDF signal, dictates that atom-pairs (PDF intensity) should be present at a larger $r$ than is seen in the experimental data; the observed intensity is transferred to lower $r$ compared to the model.

Inspection of a simulated PDF, decomposed into atom-pair specific partial contributions (Fig.~\ref{fig:misfit}(c)), discloses that the intensity in this region predominately originates from 3 distinct Ru-Ru NN contributions in the $Pnma$ structure described above.
The high-$r$ portion of this 3.1~\AA\ feature comes from the longest of these Ru-Ru NN pairs, which form the rails of the zig-zag ladder along the $b$-axis and have edge sharing RuP$_{6}$ connectivity, marked in red in Fig.~\ref{fig:struc_prop}(a) and Fig.~\ref{fig:misfit}(a, b).

\begin{figure}
\includegraphics[width=0.485\textwidth]{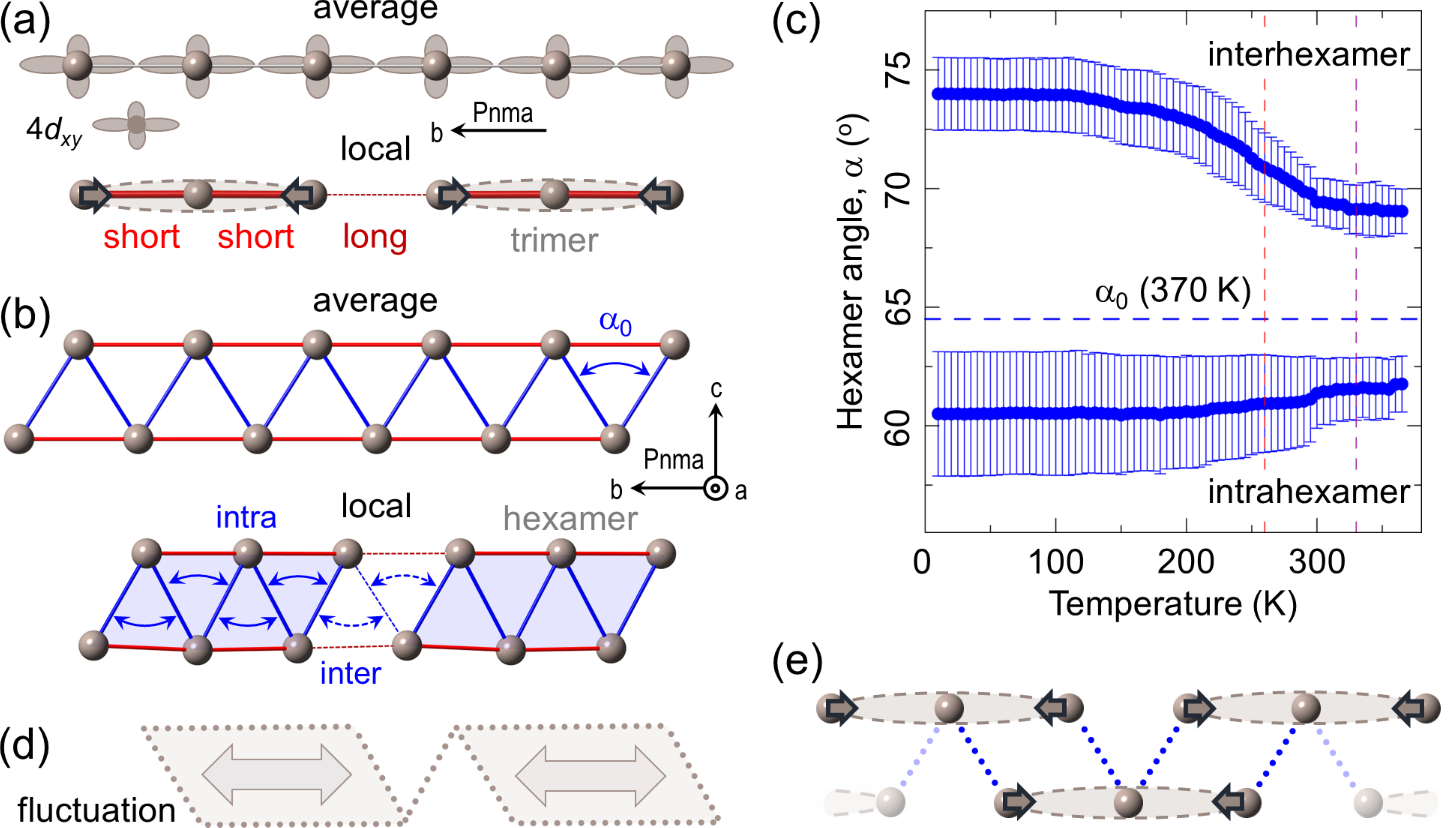}
\caption{\label{fig:hexamer}
Local distortions at 370~K.
(a) Homogeneous chains with 4$d_{xy}$ \t2g\ Ru orbitals ($Pnma$ average view, top) undergo trimerization distortions ($P2_1/c$ local view, bottom).
(b) Regular zig-zag ladders in the $bc$-plane ($Pnma$ average view, top) break into hexamer segments ($P2_1/c$ local view, bottom).
(c) Local hexamer bond angle, $\alpha$(T): symbols represent mean values, vertical bars are standard deviations.
Uniform average ($\alpha_{0}$, horizontal dashed line, $Pnma$ view), and local bimodal distribution of intrahexamer ($\alpha<\alpha_{0}$) and interhexamer ($\alpha>\alpha_{0}$) angles (2:1 relative abundance, $P2_1/c$ view).
(d) Orientational hexamer fluctuation.
(e) Non-hexamer configuration of trimers on a zig-zag ladder.
}
\end{figure}

\begin{figure*}
\includegraphics[width=0.95\textwidth]{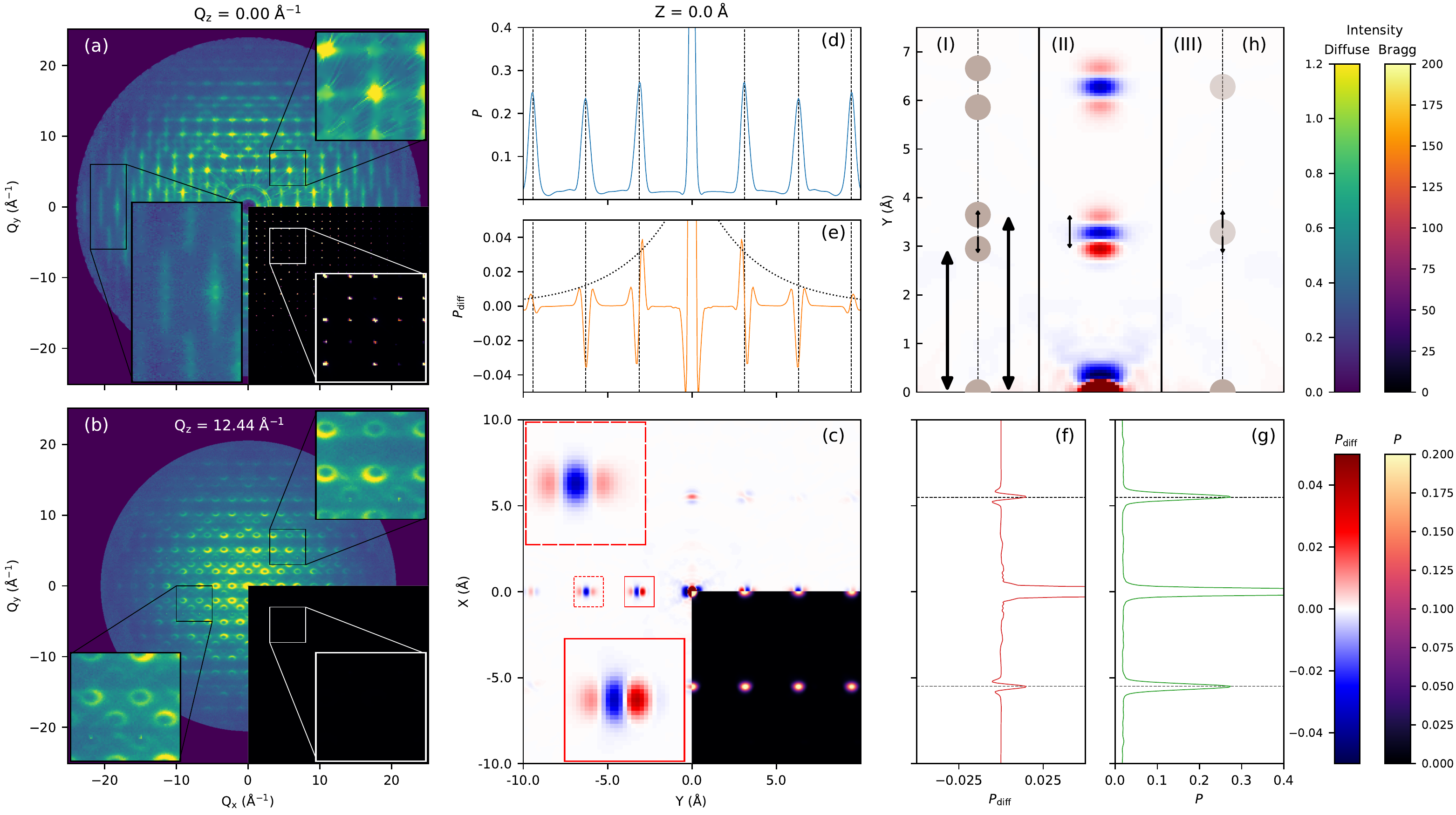}
\caption{\label{fig:3dpdf}
Single crystal view of the local state at high temperature.
Slices of the diffuse intensity distribution measured from a \rup\ single crystal at 350~K at (a) $Q_z = 0$~\RAA (containing Bragg peaks) and at (b) $Q_z = 12.44~$~\RAA (between Bragg peaks).
The lower right quadrants in (a, b) contain the full intensity distribution and are shown over a broad intensity range to highlight Bragg peaks, while other quadrants are a result of removing all Bragg peaks and shown over a narrower intensity range.
Features of interest are shown inset on an enlarged scale.
After Fourier transform, the (c) $Z = 0~$\AA\ cut of the 3D-$\Delta$PDF shows the most significant features.
The full 3D-PDF is shown in the lower right quadrant of (c) to illustrate the average position of the Ru-Ru pairs in this plane, with the Ru-Ru nearest neighbor and next-nearest neighbor features shown enlarged in the lower and upper left insets, respectively.
Line cuts for the $\Delta$ and full PDF at $Z = 0~$\AA\ along the (d, e, respectively) Y- and (f, g, respectively) X-axes are shown, with the position of Ru-Ru pairs highlighted by dotted black lines.
Dotted envelope in (e) illustrates a decay of local correlations.
The 3D-$\Delta$PDF signal along the X = Z = 0~\AA\ is consistent with the presence of trimers along this direction, as depicted in (h).
Strong (blue) negative features in (hII) indicate that Ru-Ru pairs in the local structure are displaced off the average pair distance, shown along the dashed line and shifted to the right for clarity (hIII).
Strong (red) positive features in (hII) indicate these Ru-Ru pairs are shifted to longer and shorter distances in the local structure, as represented by grey circles and marked by black arrows along the left-shifted dashed line (hI).
The local distortion moving Ru-Ru nearest neighbor pairs to shorter $r$ is about twice as likely as that moving these pairs to greater $r$, as implied by relative intensity of the features, consistent with the presence of two short and one long bond associated with trimerization.
}
\end{figure*}


The local distortion was assessed with a model utilizing $P2_1/c$ constraints, which describes the low temperature structure of RuAs~\cite{koteg;prm18}.
Fit of this broken symmetry model to the PDF data at 370~K over the range $1.75<r<15$~\AA\ results in $R_w = 9\%$, shown in Fig.~\ref{fig:misfit}(f, g).
The model adequately describes the measured PDF signal for $r<15$~\AA, successfully remedying the misfit produced by the $Pnma$ model at ca. 3.1~\AA, and decreasing $R_w$ from $23\%$ to $10\%$ for $r<3.9$~\AA.
While the $P2_1/c$ model describes the local structure well, it poorly fits the longer range data; if this model is used to compute the predicted PDF signal for $r>15$~\AA, there is significant disagreement between the model and observed PDF at 370~K (Fig.~\ref{fig:misfit}(h)) with $R_w = 34\%$.
The measured signal in the high $r$ region is visibly sharper than the $P2_1/c$ model signal, consistent with higher ($Pnma$) symmetry on longer length-scales in this temperature regime.

The $P2_1/c$ cell is related to the $Pnma$ cell by the transformation matrix
$
\begin{bmatrix}
1  & 0 & -1   \\
1  & 0 &  2  \\
0 & -3 &  0
\end{bmatrix}
$
and thus involves a 9-fold increase in unit cell volume.
The $P2_1/c$ cell contains 9 symmetry nonequivalent Ru species, and the 3 unique Ru-Ru NN pairs present in the $Pnma$ cell each split into 36 distinct Ru-Ru NN pairs in the $P2_1/c$ cell.
The temperature evolution of the local structure was quantified by fitting the PDF data for $10<T<370$~K with the $P2_1/c$ model for $1.75<r<15$~\AA.
We have then assigned each of the pairs in the refined $P2_1/c$ cell to the associated pair present in the $Pnma$ cell.
This assignment should yield 3 unique groups, each composed of 36 Ru-Ru NN pairs.
We however found that the group associated with Ru-Ru NN pairs parallel to the $Pnma$ $b$-axis (red connections in Fig.~\ref{fig:struc_prop}(a) and Fig.~\ref{fig:misfit}(a, b)) was bimodal, with 24 Ru-Ru NN pairs showing a shorter mean pair distance and 12 Ru-Ru NN pairs showing a longer mean pair distance, that is, a ratio of 2:1 of short:long pairs.

The pair distance distribution extracted from $P2_1/c$ model fits is presented in Fig.~\ref{fig:misfit}(i).
Here the plot markers represent the mean of the Ru-Ru pair distance population, with the plot error bars representing the standard deviation of this population.
The plot colors are consistent with the Ru-Ru pairs shown in Fig.~\ref{fig:struc_prop}(a) and Fig.~\ref{fig:misfit}(a-c), where the Ru-Ru NN pairs forming rails along $b$-axis, represented by red plot markers in Fig.~\ref{fig:misfit}(i), have been split into two sub-populations.
At 370~K the splitting magnitude, Fig.~\ref{fig:misfit}(i), is $\approx$ 0.4~\AA, a remarkably large distortion by crystallographic standards.
The magnitude does not change significantly across T$_{1}$, but it increases substantially below T$_{2}$, reaching 0.6~\AA\ at 100~K.

The undistorted crystallographic $Pnma$ structure at 370~K is a host to a complex local distortion, portrayed by interatomic distances and angles derived from a $P2_1/c$ model fit.
Crystallographically, the zig-zag ladders in the $bc$-planes are undistorted.
Linear Ru chains, constituting the rails along the $b$-axis in the $Pnma$ symmetry, feature uniform Ru-Ru distances (Fig.~\ref{fig:hexamer}(a), top).
The Ru-Ru rung distances bridging the ladder rails are also uniform, with all associated rung bond angles identical, $\alpha_{0}$ (Fig.~\ref{fig:hexamer}(b), top).
On the other hand, the local structure exhibits a short-short-long sequence of Ru-Ru distances along the rails (Fig.~\ref{fig:hexamer}(a) bottom).
This sequence corresponds to local nearly linear Ru trimers, a consequence of bond-charge disproportionation, implying residual covalency and, hence, charge localization above T$_1$.
Additionally, the rung bond angles from local $P2_1/c$ model form a non-uniform distribution.
These angles locally split into two well separated groups: one group of 4 adjacent angles observably smaller than the crystallographic average, $\alpha<\alpha_{0}$, and another group of 2 adjacent angles significantly larger than the average, $\alpha>\alpha_{0}$ (Fig.~\ref{fig:hexamer}(b), bottom and Fig.~\ref{fig:hexamer}(c)).
This distribution of angles means that local Ru trimers arrange into hexamers on the zig-zag ladder.
Conversely, the Ru distances constituting uniform zig-zag chains along the $Pnma$ $a$-axis have a rather narrow local distribution (0.04~\AA-0.08\AA), Fig.~\ref{fig:misfit}(i), as they do not distort appreciably at high temperature.
The dominant local distortion at T~$>$~T$_{1}$ is the Ru trimerization, with trimers further aligned into hexamers.
There is no discernible out of plane coupling, consistent with a quasi-one-dimensional disorder implicated by the ADP anisotropy.

The magnitude of local distortions changes marginally on cooling across T$_1$, whereas its dramatic increase is observed below T$_2$, as shown in Fig.~\ref{fig:misfit}(i), (j), and Fig.~\ref{fig:hexamer}(c).


%
\subsection{Local structure from single crystal PDF}
%


The observations garnered from the powder PDF analysis establish the presence of a local structure distortion and provide insights into its nature.
A complementary view of the distortion, independent of $P2_1/c$ model, is offered by a three-dimensional (3D) differential PDF analysis~\cite{weber;zk12} that probes the difference between the average and local structures directly~\cite{krogs;nm20}.
This enables exploring the presence of distortions in the high-temperature phase in a single-crystal specimen of \rup, and their 3D aspects.
For this, we have measured the 3D-$\Delta$PDF at 350~K.


The diffuse intensity distribution shows broad, rod-like features in the $Q_z=0$~\RAA\ slice, running parallel to the $Q_y$ direction,  as shown in Fig.~\ref{fig:3dpdf}(a).
The diffuse intensity is not confined to planes containing Bragg peaks, and also shows interesting features between Bragg peaks,  as can be seen in Fig.~\ref{fig:3dpdf}(b), implying short range correlations.

The corresponding 3D-$\Delta$PDF is highly anisotropic (Supplementary Note 2), with the strongest features existing in the $Z=0$~\AA\ plane, confined to line $X=0$~\AA\, as highlighted in Fig.~\ref{fig:3dpdf}(c-e).
These features are coincident with the location of overlapping Ru-Ru and P-P pair correlation vectors.
It is however likely that the dominant contribution is from Ru-Ru rather than P-P pairs, as the former have an order of magnitude stronger scattering scale (see Fig.~\ref{fig:misfit}(c)), and inspection of $\mathbf{r}$ locations where non-overlapping P-P pair correlation vectors are expected shows rather weak signal.
The locations of the strongest observed features are exactly those implicated in the powder PDF analysis, representing Ru-Ru pairs running parallel to the orthorhombic $b$-axis.
As highlighted in Fig.~\ref{fig:3dpdf}(f, g), 3D-$\Delta$PDF correlations are weak in the $Z=0$~\AA\ plane along the $X$ direction and are of the opposite sign (blue-red-blue, corresponding to negative-positive-negative) to those observed along the $Y$ direction (red-blue-red), consistent with in-phase thermal vibration of correlated motion without an observable displacive component~\cite{weber;zk12}.

\begin{figure}
\includegraphics[width=0.48\textwidth]{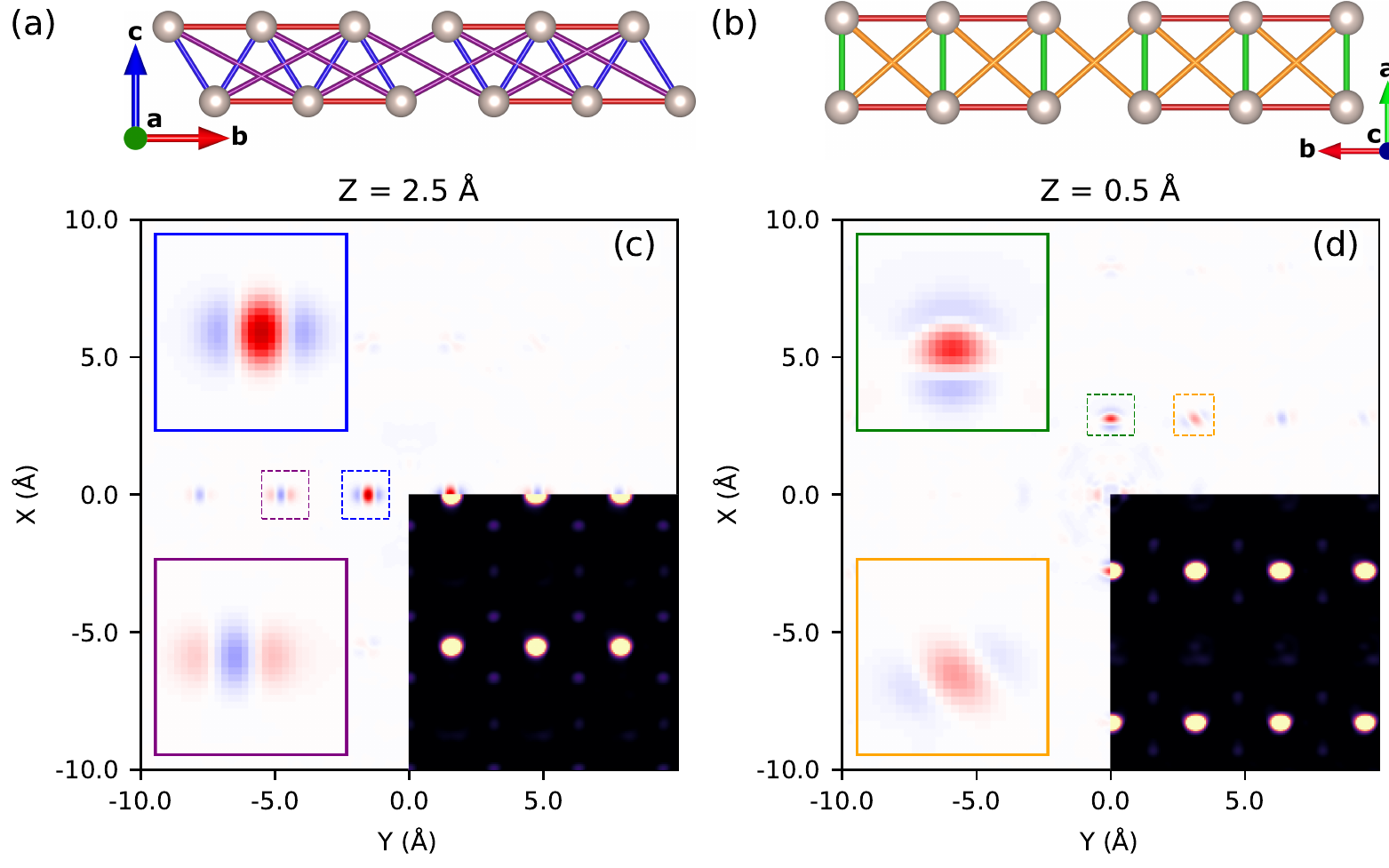}
\caption{\label{fig:3dpdf_second}
Local Ru inter-rail correlations.
Diagrams depicting Ru chains running along the $Pnma$ $b$-axis as viewed along the (a) $Pnma$ $a$-axis and (b) $Pnma$ $c$-axis.
The relevant 3D-$\Delta$PDF slices at $Z = 2.5~$\AA\ and $Z = 0.5~$\AA\ are shown in (c) and (d) respectively.
The 3D-$\Delta$PDF feature associated with Ru-Ru NN pairs connected by blue lines in (a), is highlighted in blue squares in (c), while the Ru-Ru next NN pairs connected by purple lines in (a), is highlighted in purple squares in (c).
Similarly, Ru-Ru NN and next NN pairs are marked by green and orange contacts, respectively, in (b) and highlighted in squares of matching color in (d).
Total 3D-PDF, shown in the lower right quadrant of (c) and (d), illustrates the average location of the Ru-Ru pair vectors in these planes.
}
\end{figure}


The signal associated with Ru-Ru correlations in the $Z=0$~\AA\ plane along the line $X=0$~\AA\ is rather unique.
The first feature at $Y= 3.1$~\AA\ contains a central negative (blue) lobe, with two outer positive (red) lobes (see e.g. Fig.~\ref{fig:3dpdf}(h)), indicating that Ru-Ru nearest neighbors, marked by red bonds throughout this work, are displaced off the average pair position, along the orthorhombic $b$-axis.
This is consistent with bimodal distribution of Ru-Ru interatomic vectors in this direction.
Importantly, the positive lobe closer to the origin is about twice as strong as the positive lobe further from the origin, suggesting that Ru-Ru nearest neighbors in the local structure adopt a 2:1 ratio of short to long bonds along the orthorhombic $b$-axis, indicative of a Ru trimer and consistent with powder PDF observations.



The 3D-$\Delta$PDF also provides insights on rail-to-rail correlations within the zig-zag ladder in the orthorhombic $bc$-plane.
The $Z=2.5$~\AA\ plane along the line $X=0$~\AA\, shown in Fig.~\ref{fig:3dpdf_second}(c), gives information on neighboring Ru chains along the orthorhombic $c$-axis, as shown in Fig.~\ref{fig:3dpdf_second}(a).
The NN Ru-Ru pairs along this direction, highlighted in Fig.~\ref{fig:3dpdf_second}(a) by blue connections, do not show a distortion off the average structure site, but show a positive $\Delta$PDF lobe at the location of the average structure NN Ru-Ru pair, the signature of strong bonding.
Conversely, Ru-Ru next NN pairs along this direction, highlighted in Fig.~\ref{fig:3dpdf_second}(a) by purple connections, do exhibit a displacive distortion off the average structure site.
This indicates existence of shorter and longer next NN inter-rail pairs and supports the powder PDF observation of a hexamer-like association of trimers on the ladder.

The $Z=0.5$~\AA\ plane, shown in Fig.~\ref{fig:3dpdf_second}(d), gives information on Ru-Ru correlations in the orthorhombic $ab$-plane, as shown in Fig.~\ref{fig:3dpdf_second}(b).
The NN Ru-Ru pairs in this plane, highlighted in Fig.~\ref{fig:3dpdf_second}(b) by green connections, do not show a distortion off the average structure site, but show
a positive $\Delta$PDF lobe at the location of the average structure NN Ru-Ru pair, reflecting in-phase thermal motion, associated with strong bonding.
Similarly, Ru-Ru next NN pairs in this plane, highlighted in Fig.~\ref{fig:3dpdf_second}(b) by orange connections, also do not show a displacive distortion off the average structure site.
This, taken together with the $P2_1/c$ model derived powder PDF observations of rather small distortions of the green zig-zag chains along the orthorhombic $a$-axis, suggests
weak inter-ladder hexamer correlations.

Deviations from the average $Pnma$ structure are not detected beyond 15~\AA\ in powder PDF analysis, whereas the signatures of displacive distortions in 3D-$\Delta$PDF are highly anisotropic and confined to the orthorhombic $bc$-plane over a similarly narrow spatial range.
This corroborates quasi-one-dimensional character of fluctuations correlated over a length-scale of about two hexamer units.

\section{Discussion and Conclusions}
\label{section:dis}
\subsection{Ru$_{6}$ hexamers - origin and fluctuations}
%

\begin{figure}
\includegraphics[width=0.485\textwidth]{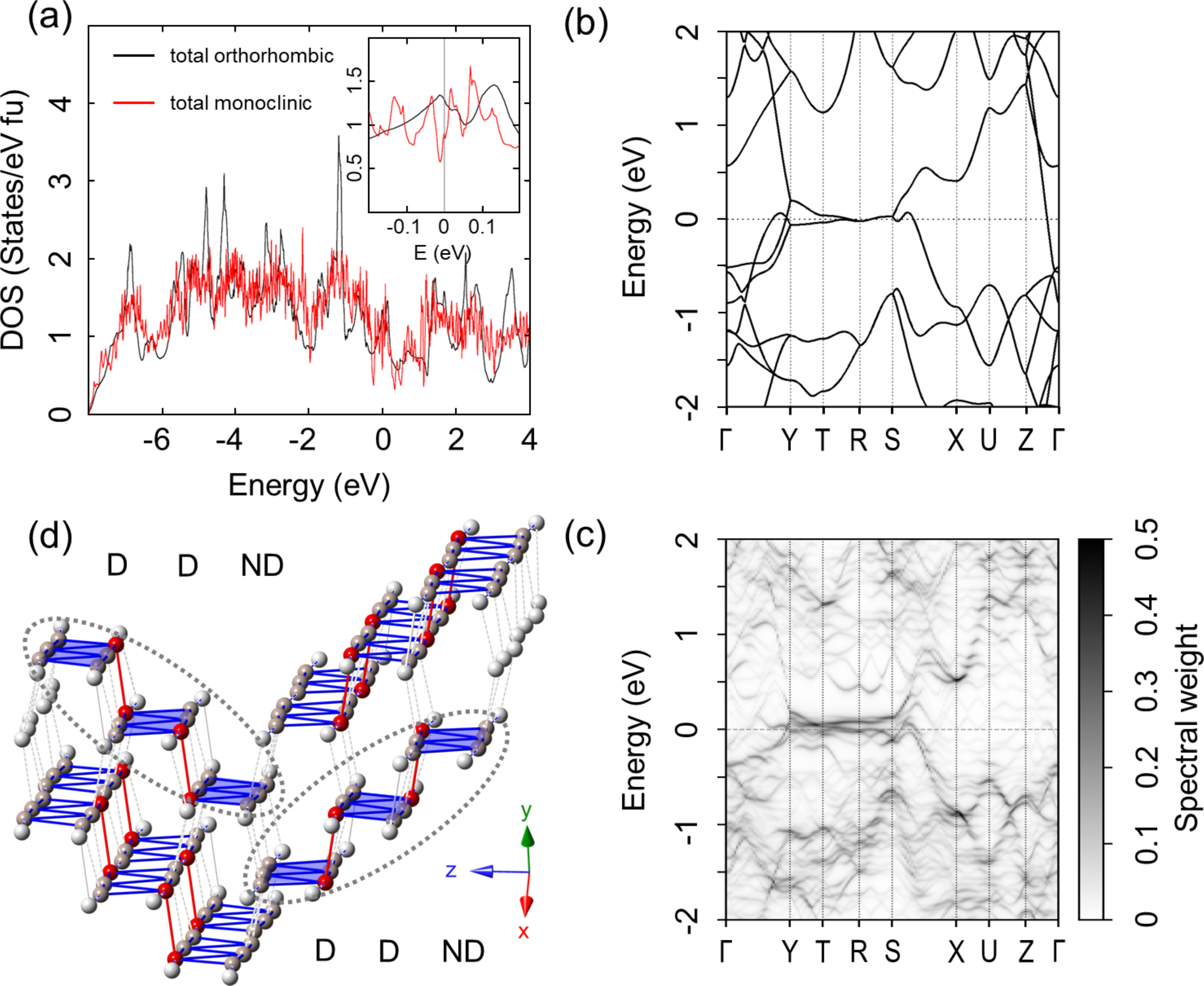}
\caption{\label{fig:calculation}
Electronic structure and Ru oligomerization.
(a) Comparison of DOS calculated for the orthorhombic and monoclinic structures derived from 370~K PDF analysis.
Inset: DOS near E$_{f}$ reveals $\sim$100 meV splitting.
Unfolded band structure in the Brillouin zone of the orthorhombic cell for (b) undistorted orthorhombic structure expressed in the monoclinic setting and (c) distorted monoclinic structure, where the line intensity reflects spectral weight.
In (c), a small gap opens along the direction of the flat bands (T-R-S), consistent with the splitting of DOS in (a).
(d) Segment of distorted structure from 15~\AA\ fit to 10~K PDF data.
The zig-zag chains, undistorted in the orthorhombic structure, form triplets comprised of two chain types.
The first type, D, features shorter ($\sim$2.79~\AA) dimer-like Ru-Ru contacts (colored red) connecting three Ru$_{6}$ hexamers on adjacent ladders, and longer Ru-Ru contact ($\sim$2.85~\AA\ on average).
The second type, ND, is less distorted and with no dimer-like contacts.
The triplet consists of two D-type and one ND-type chains.
Hexamers organize into higher-order oligomers, implying modified orbital occupancy at T$_{2}$ as inter-ladder correlations develop.
}
\end{figure}

The driving force behind this peculiar local oligomer - hexamer built of trimers is the Fermi surface instability imposed by $Pnma$ symmetry and nominal $4d^{5}$ filling, embodied in flat degenerate bands of Ru 4$d_{xy}$ \t2g\ orbital character and DOS peaked at E$_f$~\cite{koteg;prm18,ootsu;prb20,goto;pp15}.
Simultaneous contraction of adjacent Ru pair distances sharing central Ru results in the observed sizeable distance bifurcation.
Due to edge-sharing of RuP$_{6}$ along the $b$-axis, Fig.~\ref{fig:struc_prop}(a), the distortions involve precisely \t2g\ $xy$ orbital overlaps, Fig.~\ref{fig:hexamer}(a).
Large distortions of the longest of the three distinct NN Ru pair distances necessitate bond-charge disproportionation and increased covalency of the short Ru-Ru contacts, pointing to their electronic origin.
Density functional theory calculations on fully converged average and local models of the 370~K PDF data support this picture.
Calculations using the orthorhombic average structure reproduce the instability - flat bands are seen in the T-R (along the $Pnma$ a-axis) and R-S (along the $Pnma$ c-axis, Supplementary Note 6) directions, Fig.~\ref{fig:calculation}(b), with peak in DOS at E$_f$, black trace in Fig.~\ref{fig:calculation}(a).
In contrast, calculations using global monoclinic distorted phase based on the local model show removal of the instability, Fig.~\ref{fig:calculation}(c), by opening a small gap around the T-R-S line and splitting of DOS at E$_f$ (red trace in Fig.~\ref{fig:calculation}(a)) thus lifting the degeneracy of the flat bands, albeit partially.
While further theoretical investigations are necessary to fully comprehend the driving mechanism behind the monoclinic distortion and metal to insulator transition, our calculations confirm casuality between the electronic instability and crystallographic symmetry breaking.
The distortions driven by this instability could then be considered as a fingerprint of PG fluctuations.

In this picture the trimerization represents a primary response, with the hexamers being, arguably, a consequence of elastic energy penalty minimization associated with short range trimer correlations.
To illustrate this, we consider a non-hexamer arrangement of trimers on a ladder, Fig.~\ref{fig:hexamer}(e), where the trimers on the opposite rails are offset by one Ru position with respect to their hexamer arrangement depicted in Fig.~\ref{fig:hexamer}(b).
Here, central Ru atom of the trimer on the bottom rail faces the long inter-trimer distance on the top rail.
Compared to hexamer configuration, this configuration involves a larger number of energetically expensive rung bond-stretching distortions, dotted blue lines in Fig.~\ref{fig:hexamer}(e).
Since the trimer pairing minimizes bond-stretching and engages fewer energetically less expensive bond-bending distortion modes, hexamer formation is favored.

The undistorted structure is retrieved by spatiotemporal averaging of the distortions.
As an energy-integrating method PDF does not distinguish static from dynamic distortions.
While purely static distortions, incoherently distributed across the ladders and randomized by defects would be consistent with lack of long range order, local hexamers are more likely dynamic, brought about by temporal fluctuations of the trimer bond-charge.
Dynamic alternations of hexamer orientations, illustrated in Fig.~\ref{fig:hexamer}(d), are expected to be entropically stabilized.

\subsection{Implications for electronic properties}

The observed Ru$_{6}$ fluctuations provide a rationale for poor metallic conduction of \rup\ above the two-stage transition, seen previously in some metallic spinels~\cite{bozin;nc19,yang;prb20}.
They further act as a precursor to the transitions, evident from the evolution of local distortion with temperature, Figs.~\ref{fig:misfit}(i) and~\ref{fig:hexamer}(c).

The lack of local structure changes at T$_{1}$ implies the PG transition has order-disorder character (see Supplementary Note 4), where hexamer stripes are gradually formed, governed by an energy scale lower than that of hexamer formation.
Inter-hexamer correlations grow in the PG phase on approaching the NM transition, coincident with antiferromagnetic fluctuations seen by nuclear magnetic resonance~\cite{li;prb17,kuwat;jpsj18}.
The intra-hexamer bonding strengthens, reflected in contraction of blue (rungs) and short red (intra-trimer) Ru-Ru distances, Fig.~\ref{fig:misfit}(i), whereas the ladder inter-hexamer coupling weakens, seen as relaxation of inter-trimer contacts (long red distances in Fig.~\ref{fig:misfit}(i)) and inter-hexamer angles (top branch in Fig.~\ref{fig:hexamer}(c)).
Notably, the distribution of Ru-Ru contacts along the orthorhombic $a$-axis (zigzag chains, green data in Fig.~\ref{fig:misfit}(i)) broadens, implying development of 3D hexamer correlations.

Below T$_{2}$ the local structure changes more abruptly, and the bimodal splitting, Fig.~\ref{fig:misfit}(j), increases by 50~\%\ as hexamers compact and distance from each other on the ladders.
Such large local distortion enhancements were found in systems with spin singlet ground states exhibiting orbital molecules~\cite{bozin;nc19,yang;prb20,koch;prl21}.
Interestingly, fits of $Pnma$ and $P2_1/c$ models to the diffraction data, although both strictly inadequate, indicate negative thermal expansion of respective ($a$, $b$) and ($a$, $c$) axes in the PG phase (Supplementary Note 3).
Similar negative thermal expansion was seen in CrSe$_{2}$~\cite{brugg;pb80} in association with metal cluster fluctuations, where Peierls-like instability leads to an orbitally ordered ground state of linear trimers~\cite{kobay;ic19}.
While detailed crystallographic characterization is required to grasp full complexity of the ground state, assessment of local structure from 10~K PDF fit suggests further complex oligomerization such as depicted in Fig.~\ref{fig:calculation}(d).
The structure features two types of zigzag chains consistent with cell tripling in all three directions (see Supplementary Note 5).
Two neighboring chains of one type (D in Fig.~\ref{fig:calculation}(d)) are distorted and carry short dimer-like inter-hexamer contacts, whereas the third chain constituting the other type (ND in Fig.~\ref{fig:calculation}(d)) does not exhibit such short contacts.
This results in triplets of dimer-bound hexamers.

Local hexamers observed by PDF, as fluctuations of the PG phase, are an important component of the observed complex phase diagram of binary Rh$Pn$.
The SC state in both RuP and RuAs derived compositions with a maximum transition temperature, at the quantum critical point of the PG phase, was attributed to the conventional single-band weak coupling Bardeen-Cooper-Schrieffer (BCS) model~\cite{anand;prb18}.
Notably, a linear temperature dependence of the upper critical field, H$_{c2}$, persists down to the lowest temperature with no saturation tendency, a behavior not expected for an isotropic single-band BCS superconductor.
This observation as well as dramatically different SC dome maximum transition temperatures in RuP, RuAs and RuSb families, raise the question of the role of Ru$_{6}$ fluctuations in SC pairing~\cite{taill;arcmp10} and enhancement of T$_{c}$, e.g. via tuning of the effective electron-hole interactions~\cite{basov;np11}, near the PG quantum critical point~\cite{badou;n16} and the interplay with reported antiferromagnetic fluctuations in the PG regime.

\section{Methods}
\label{section:meth}
{\it Sample preparation and characterization.}
RuP crystals were grown from excess Sn~\cite{fisk;b;hbpcre89}.
The magnetization was measured in a Quantum Design MPMS-XL5.
\medskip

{\it Powder diffraction experiment.}
High energy synchrotron radiation powder total scattering experiments were conducted at the 28-ID-1 (PDF) beamline at the National Synchrotron Light Source-II (NSLS-II) at Brookhaven National Laboratory (BNL).
The sample was loaded into a 1~mm inner diameter kapton capillary and data collected from 370~K to 10~K in 5~K steps on cooling using a liquid He$_2$ cryostat.
Measurements were carried out in the rapid acquisition pair distribution function (RaPDF) mode~\cite{chupa;jac03}, with an x-ray energy of 74.47~keV ($\lambda = 0.1665$~\AA).
A two-dimensional (2D) PerkinElmer area detector was used, with a sample-to-detector distance of 204~mm for the PDF measurements and 1007~mm for higher resolution powder diffraction measurements, both determined by calibrating to a sample of known lattice parameter (Ni).

The 2D data were integrated and converted to intensity as a function of momentum transfer $Q$ using the software \pyfai~\cite{Ashiotisfastazimuthalintegration2015a}.
The program \pdfgetxthree v2.0~\cite{juhas;jac13} was used to correct, normalize, and Fourier transform the diffraction data to obtain the experimental xPDF, $G(r)$, up to a momentum transfer of $\qmax=27~\ia$ which was chosen as the best tradeoff between real-space resolution and noise in the data.

\medskip

{\it Structure modeling.} The PDF modeling programs \pdfgui and \cmi were used for powder PDF structure refinements~\cite{farro;jpcm07,juhas;aca15}. In these refinements $U_{iso}$~(\AA$^2$) is the isotropic atomic displacement parameter (ADP) and the ADPs of the same type of atoms are constrained to be the same; $\delta_2$~(\AA$^2$) is a parameter that describes correlated atomic motions~\cite{jeong;jpc99}.
The PDF instrument parameters $Q_{damp}$ and $Q_{broad}$ determined by fitting the PDF from the well crystallized standard sample under the same experimental conditions are fixed in the structural refinements on \rup\ dataset.

\medskip

{\it The Pearson correlation coefficient.} For comparing powder PDF data at distinct temperature points, the Pearson correlation coefficient, $r_{T_1T_2}$ was used, defined as
\begin{equation}
r_{T_1T_2} = \frac{\sum (G_{T_1}-\mu_{T_1})(G_{T_2}-\mu_{T_2})}{\sqrt{\sum(G_{T_1}-\mu_{T_1})^2\sum(G_{T_2}-\mu_{T_2})^2}}
\end{equation}
where $G_{T_1}$ and $G_{T_2}$ are the PDF data points for temperatures 1 and 2, respectively, and $\mu_{T_1}$ and $\mu_{T_2}$ are mean of the PDF data points for temperatures 1 and 2. respectively.

{\it Single crystal experiment and 3D-$\Delta$PDF analysis.} High energy synchrotron radiation single crystal total scattering experiments were conducted at the P21.1 beamline at the Positron-Elektron-Tandem-Ring-Anlage (PETRA III) facility at Deutsches Elektronen-Synchrotron (DESY), using an x-ray beam of 101.63 keV energy ($\lambda$ = 0.1220~\AA) sized to $0.5\times0.5$~mm$^2$.

The crystal was mounted on a ca. 2~mm carbon fiber with GE varnish, with the crystal carefully aligned to achieve adequate centering.
Measurements were carried out in a Displex closed loop helium cryochamber under vacuum at 350~K.
Each diffraction image was collected with a PILATUS3X 2M CdTe area detector located 500~mm from the sample.
Data were collected with two distinct beam attenuations to increase dynamic range and 8 distinct detector positions to fill detector module gaps and increase the maximum momentum transfer $\qmax$.
Detector distance, tilt, and rotation for each detector position were calibrated using a CeO$_2$ standard measured in an identical geometry.
With this beam energy and geometry, the detector provides a 21~\ia\ range coverage of reciprocal space ($\qmax=21$~\ia), giving $\Delta_{r}=0.3$~\AA.

Data were collected while the crystal was exposed to x-rays under continuous rotation over 200 degrees, with an exposure speed 0.1 frame/second and rotation rate of 0.1 degrees/second (2,000 total frames per measurement).
The 2D detector images were transformed to intensity as a function of momentum transfer $\mathbf{Q}$ on a $801 \times 801 \times 801$ grid (0.031~\RAA step size) using custom built software, up to a momentum transfer of $\mathbf{Q}_{\mathrm{max}}=21~\ia$ which was chosen as the best tradeoff between real-space resolution and noise in the data.
Data were symmetry averaged and reciprocal space Bragg peaks were removed according to the $Pnma$ space group with a spherical punch of diameter 15 voxels (0.47~\RAA), and 3D-$\Delta$PDF produced by Fourier transform as described in previous work~\cite{koch:aca21}.

\medskip

{\it Density functional theory calculations.} The calculations were done by using Vienna Ab Initio Simulation Package~\cite{VASP1,PAW}.
Perdew-Burke-Ernzerhof exchange-correlation functionals~\cite{PBE} within the generalized gradient approximation were used in all the calculations.
An appropriate energy cut-off of 500~eV and Monkhost-Pack k-mesh with a grid spacing less than 2$\pi\times$0.03 \AA$^{-1}$ was adopted in our calculations.
Band unfolding of the monoclinic distorted phase into the undistorted orthorhombic phase was done by using the method of Popescu and Zunger~\cite{PopescuZunger2012} as implemented in the PyVASPwfc software package~\cite{pyvaspwfc}.

\medskip

%
%
%
\section{Acknowledgments}
We gratefully acknowledge Olof Gutowski for assistance with the single crystal measurements, and Joe D. Thompson for collecting the magnetic susceptibility data.
Work at Brookhaven National Laboratory was supported by U.S. Department of Energy, Office of Science, Office of Basic Energy Sciences (DOE-BES) under contract No. DE-SC0012704.
\rjkadd{We acknowledge DESY (Hamburg, Germany), a member of the Helmholtz Association HGF, for the provision of experimental facilities. Parts of this research were carried out at beamline P21.1 at PETRA III.}
ESB acknowledges the Stephenson Distinguished Visitor Programme for supporting his stay at DESY in Hamburg.
This research was supported in part through the Maxwell computational resources operated at Deutsches Elektronen-Synchrotron DESY, Hamburg, Germany.
Work at Los Alamos National Laboratory was performed under the auspices of the US Department of Energy, Office of Basic Energy Sciences, and Division of Materials Sciences and Engineering under project "Quantum Fluctuations in Narrow Band Systems".


\vfill\newpage

\renewcommand\thefigure{S\arabic{figure}}
\setcounter{figure}{0}

\end{document}